\documentclass[aps,preprint]{revtex4}%
\usepackage{amsfonts}
\usepackage{amsmath}
\usepackage{amssymb}
\usepackage{graphicx}%
\begin{document}
\preprint{CTP-SCU/2013001}
\title{Fermion's tunnelling with effects of quantum gravity}
\author{Deyou Chen}
\email{dchen@cwnu.edu.cn}
\affiliation{Institute of Theoretical Physics, China West Normal University, Nanchong, 637009, China}
\author{Houwen Wu}
\email{iverwu@uestc.edu.cn}
\affiliation{Center for Theoretical Physics, College of Physical Science and Technology,
Sichuan University, Chengdu, 610064, China}
\author{Haitang Yang}
\email{hyanga@scu.edu.cn}
\affiliation{Center for Theoretical Physics, College of Physical Science and Technology,
Sichuan University, Chengdu, 610064, China}

\begin{abstract}
{In this paper, using
Hamilton-Jacobi method,  we address the tunnelling of fermions in
a 4-dimensional Schwarzschild spacetime. Base on the generalized
uncertainty principle, we introduce the influence of quantum
gravity. After solving the equation of motion of the spin $1/2$
field, we derive the corrected Hawking temperature. It turns out
that the correction depends not only on the black hole's mass but
also on the mass (energy) of emitted fermions. It is of interest
that, in our calculation, the quantum gravity correction
decelerates the temperature increase during the radiation
explicitly. This observation then naturally leads to the remnants
in black hole evaporation. Our calculation shows that the residue
mass is $\gtrsim M_p/\beta_0$, where $M_p$ is the Planck mass and
$\beta_0$ is a dimensionless parameter accounting for quantum
gravity effects. The evaporation singularity is then avoided.}

\end{abstract}
\maketitle
\tableofcontents

\section{Introduction}

Hawking radiation is described as a quantum tunnelling effects of
particles at  horizons of black holes
\cite{SWH,KW,PW,ZZ,JWC,KSAE}. With the consideration of the
background variation in black hole evaporation, Parikh and Wilczek
studied the tunnelling behaviors of massless scalar particles
\cite{PW}. They derived the modified emission spectra for
spherically symmetric black holes. The leading corrections to the
Hawking temperature are found to be also dependent on the energy
of emitted particles. This work was extended to massive and
charged scalar particles. The Hawking radiation of general black
holes was studied \cite{ZZ,JWC}. For an outgoing massive particle,
the equation of motion is different from that of a massless
particle. The trajectory of massless particles is a null geodesic.
While the massive particle's motion satisfies de Broglie wave and
is the phase velocity of outgoing particles. Subsequently, the
tunnelling behaviors of fermions were carefully investigated with
Hamilton-Jacobi method by Kerner and Mann \cite{KM}. The Hawking temperatures
were recovered by the fermions tunnelling. The extension
of this work to complicated spacetimes are referred to
\cite{LR,CJZ,QQJ,HCNVZ,LY}.

Various theories of quantum gravity predict the existence of a
minimum  measurable length \cite{PKT,ACV,KPP,LJG,GAC}. This length
can be approached from the generalized uncertainty principle
(GUP). Through the modified fundamental commutation relation
\cite{KMM}

\begin{equation}
[x_i,p_j] = i\hbar \delta_{ij}[1+\beta p^2],
\label{eq1}
\end{equation}

\noindent the expression of GUP is derived as $\Delta x \Delta
p\geq \frac{\hbar}{2}  [1 + \beta (\Delta p)^2]$, where $\beta
=\beta_0/M_p^2$. $M_p$ is the Planck mass. $\beta_0 $ is a
dimensionless parameter. From simple electroweak consideration, it
is readily to find an upper limit $\beta_0 < 10^{34}$. $x_i$ and
$p_i$ are defined by $x_i=x_{0i}$ and $p_i = p_{0i }(1 + \beta
p^2)$, respectively. $x_{0i}$ and $p_{0j}$ satisfy the canonical
commutation relations $[x_{0i},p_{0j}] = i\hbar \delta_{ij} $. The
modification of the fundamental commutation relation is not
unique. Other modifications are referred to \cite{AK,FB,ADV} and
references therein.

These modifications are widely applied to gain some information
about the  quantum properties of gravity. Black holes are effective modes to
explore effects of quantum gravity. Incorporating effects of
quantum gravity into black hole physics by GUP, some interesting
implications and results were achieved
\cite{KSY,NM,XW,BG,NS,AFA,BM}. It showed in \cite{KSY} that a
small black hole is unstable. Moreover, the constraint for a large
black hole comparable to the size of the cavity in connection with
the critical mass is needed. The characteristic size in the
absorption process, represented by the black hole irreducible
mass, was gotten in \cite{XW}. The remnant mass and corrections to
the area law and heat capacity were obtained in \cite{BG}. In
\cite{NS}, following Parikh-Wilczek tunnelling method, based on
GUP, the radiation of massless scalar particles in the
Schwarzschild black hole was discussed. The commutation relation
between the radial coordinate and the conjugate momentum are
modified with GUP. The authors treat the natural cutoffs as a
minimal length, a minimal momentum and a maximal momentum. They
addressed the tunnelling rate of black holes. The corrected
Hawking temperature was obtained and related to the energy of
emitted particles.

The purpose of this paper is to investigate fermions' tunnelling
behavior  cross the event horizon of a $4$-dimensional
Schwarzschild black hole, where effects of quantum gravity are
taken into account. We first modify the Dirac equation in curved
spacetime to reflect the influence of quantum gravity. The model
we adopt is the generalized uncertainty principle. We use the
Hamilton-Jacob method to solve the equation of motion of the
spinor field. Then the tunnelling rate and Hawking temperature are
calculated. Our results show that the quantum correction to the
Hawking temperature is dependent not only on the black hole's mass
but also on the mass and energy of emitted fermions. Moreover, the
correction slows down the temperature increase during the
evaporation. This in turn leads to the remnants in black hole
evaporation and prevents the existence of the thermodynamics
singularity.

The rest is organized as follows. In Sect.2, taking into account
effects of quantum gravity, we modify Dirac equation in curved
spacetime by GUP and get a generalized Dirac equation. In Sect.3,
the fermion tunnelling behavior in the Schwarzschild black hole is
addressed and the corrected Hawking temperature is derived. Sect.4 is devoted to our discussion and conclusion.

\section{Generalized Dirac equation in curved spacetime}

To take into account the effects of quantum gravity, we adopt the
generalized commutation relation in \cite{KMM} to modify the Dirac
equation. In Eq. (\ref{eq1}), the momentum operators are defined
by

\begin{equation}
p_i = p_{0i} (1 + \beta p^2). \label{eq2}
\end{equation}

\noindent The square of momentum operators is

\begin{eqnarray}
p^2 &=& p_i p^i = -\hbar^2 \left[ {1 - \beta \hbar^2 \left( {\partial _j \partial ^j} \right)}
\right]\partial _i \cdot \left[ {1 - \beta \hbar^2 \left( {\partial ^j\partial _j } \right)}
\right]\partial ^i\nonumber \\
&\simeq & - \hbar ^2\left[ {\partial _i \partial ^i - 2\beta \hbar
^2 \left( {\partial ^j\partial _j } \right)\left( {\partial
^i\partial _i } \right)} \right], \label{eq3}
\end{eqnarray}

\noindent where in the last step, we only keep the leading order
term of $\beta $. To account for the effects from quantum gravity,
the frequency is generalized as \cite{Hossenfelder:2003jz}

\begin{equation}
\bar \omega = E( 1 - \beta E^2), \label{eq4}
\end{equation}

\noindent with the energy operator $ E = i \hbar \partial _0 $.
Substituting  the mass shell condition $ p^2 + m^2 = E^2 $, we get
the generalized expression of energy \cite{NK,HBHRSS}

\begin{equation}
\bar E = E[ 1 - \beta (p^2 + m^2)]. \label{eq5}
\end{equation}

\noindent The tunnelling of massless scalar particles in the
Schwarzschild  black hole was studied in detail and the corrected
Hawking temperature was derived in \cite{NS}. On the other hand,
Dirac equation with the consequence of GUP in flat spacetime has
been investigated in \cite{NK}.

We start with the Dirac equation in curved spacetime,
\begin{equation}
i\gamma ^\mu \left( {\partial _\mu + \Omega _\mu } \right)\psi +
\frac{m}{\hbar }\psi = 0, \hspace{7mm} \Omega _\mu \equiv
\frac{i}{2}\omega _\mu\, ^{a b} \Sigma_{ab},  \label{eq6}
\end{equation}

\noindent where $\omega _\mu\, ^{ab}$ is the spin connection
defined  by the tetrad $e^\lambda\,_b$ and ordinary connection
\begin{equation}
\omega_\mu\,^a\,_b=e_\nu\,^a e^\lambda\,_b \Gamma^\nu_{\mu\lambda}
-e^\lambda\,_b\partial_\mu e_\lambda\,^a.
\end{equation}
The Latin indices live in the flat metric $\eta_{ab}$ while Greek
indices are raised and lowered by the curved metric $g_{\mu\nu}$.
The tetrad can be constructed from
\begin{equation}
g_{\mu\nu}= e_\mu\,^a e_\nu\,^b \eta_{ab},\hspace{5mm} \eta_{ab}=
g_{\mu\nu} e^\mu\,_a e^\nu\,_b, \hspace{5mm} e^\mu\,_a e_\nu\,^a=
\delta^\mu_\nu, \hspace{5mm} e^\mu\,_a e_\mu\,^b = \delta_a^b.
\label{eq6-1}
\end{equation}
Back in equation (\ref{eq6}), $\Sigma_{ab}$'s are the Lorentz
spinor generators defined by
\begin{equation}
\Sigma_{ab}= \frac{i}{4}\left[ {\gamma ^a ,\gamma^b} \right],
\hspace{5mm} \{\gamma ^a ,\gamma^b\}= 2\eta^{ab}. \label{eq6-2}
\end{equation}
Then one can construct the $\gamma^\mu$'s in curved spacetime as
\begin{equation}
\gamma^\mu = e^\mu\,_a \gamma^a, \hspace{7mm} \left\{ {\gamma ^\mu
,\gamma ^\nu } \right\} = 2g^{\mu \nu }. \label{eq6-3}
\end{equation}
To get the generalized Dirac equation in curved spacetime, we
rewrite eq. (\ref{eq6}) as

\begin{equation}
 - i\gamma ^0\partial _0 \psi = \left( {i\gamma ^i\partial _i + i\gamma ^\mu
\Omega _\mu + \frac{m}{\hbar }} \right)\psi , \label{eq7}
\end{equation}

\noindent where $i = 1,2,\cdots$ denotes the spatial coordinates.
The left hand-side of the equation above is related to the energy.
Using the generalized expression of energy eqn. (\ref{eq5}) and
the square of momentum operators eqn. (\ref{eq3}), only keeping
the leading order term of $\beta$, we get

\begin{eqnarray}
 - i\gamma ^0\partial _0 \psi = \left( {i\gamma ^i\partial _i +
 i\gamma ^\mu \Omega _\mu + \frac{m}{\hbar }} \right)\left( {1 +
 \beta \hbar ^2\partial _j\partial ^j - \beta m^2} \right)\psi.
 \label{eq8}
\end{eqnarray}

\noindent Therefore, the generalized Dirac equation in curved
spacetime can be written as

\begin{eqnarray}
\left[i\gamma^{0}\partial_{0}+i\gamma^{i}\partial_{i}
\left(1-\beta m^{2} \right)+i\gamma^{i}
\beta\hbar^{2}\left(\partial_{j}\partial^{j}\right)
\partial_{i}+\frac{m}{\hbar}\left(1+\beta\hbar^{2}\partial_{j}\partial^{j}
-\beta m^{2}\right)\right.\nonumber \\
\left.+i\gamma^{\mu}\Omega_{\mu}\left(1+\beta\hbar^{2}\partial_{j}
\partial^{j} -\beta m^{2}\right)\right]\psi = 0. \label{eq9}
\end{eqnarray}
\noindent This is the equation we are going to solve in the next
section.

\section{Fermion tunnelling with effects of quantum gravity}

In this section, we address the tunnelling behavior of spin-$1/2$
fermions  across the event horizon of the Schwarzschild black
hole. Effects of quantum gravity are taken into account. The
metric is given by

\begin{equation}
ds^2 = -f(r)dt^2 + \frac{1}{g(r)}dr^2 + r^2 (d\theta^2 +
\sin^2\theta d\phi^2), \label{eq10}
\end{equation}

\noindent with $ f\left(r\right)=g\left(r\right)=1-\frac{2M}{r}$,
and  $ M $ is the black hole's mass. We have set $G=c=1$. The
event horizon is located at $ r_h = 2M $. The fermion's motion is
determined by the generalized Dirac equation (\ref{eq9}). For a
spin-1/2 particle, there are two states corresponding respectively
to spin up and spin down. Follow the standard ansatz, to describe
the motion semi-classically, we assume the wave function of the
spin up state as

\begin{equation}
\Psi=\left(\begin{array}{c}
A\\
0\\
B\\
0
\end{array}\right)\exp\left(\frac{i}{\hbar}I
\left(t,r,\theta , \phi \right)\right),
\label{eq11}
\end{equation}

\noindent where $A, B$ and $I$ are functions of coordinates $ t,
r,  \theta , \phi$, and $I$ is the action of the emitted fermions.
The process of spin down is  the same as that of spin up. To solve
eqn. (\ref{eq9}), one should choose appropriate gamma matrices by
exploiting eqn.s (\ref{eq6-1})-(\ref{eq6-3}). It is
straightforward to guess a tetrad for the metric (\ref{eq10})

\[e_\mu\,^a = \rm{diag}\left(\sqrt f, 1/\sqrt g, r,
r\sin\theta \right).\]

\noindent Then, our gamma matrices are given by

\begin{eqnarray}
\gamma ^t &=& \frac{1}{\sqrt {f\left( r \right)} }\left(
{{\begin{array}{*{20}c}
 i \hfill & 0 \hfill \\
 0 \hfill & { - i} \hfill \\
\end{array} }} \right),
\quad \gamma ^\theta = \sqrt {g^{\theta \theta}}\left(
{{\begin{array}{*{20}c}
 0 \hfill & {\sigma ^1} \hfill \\
 {\sigma ^1} \hfill & 0 \hfill \\
\end{array} }} \right),\nonumber\\
\label{eq12} \gamma ^r &=& \sqrt {g\left( r \right)} \left(
{{\begin{array}{*{20}c}
 0 \hfill & {\sigma ^3} \hfill \\
 {\sigma ^3} \hfill & 0 \hfill \\
\end{array} }} \right),
\quad \gamma ^\phi = \sqrt {g^{\phi \phi}}\left(
{{\begin{array}{*{20}c}
 0 \hfill & {\sigma ^2} \hfill \\
 {\sigma ^2} \hfill & 0 \hfill \\
\end{array} }} \right),
\label{eq12}
\end{eqnarray}

\noindent with  $\sqrt {g^{\theta \theta}}= \frac {1}{r}$,  $\sqrt
{g^{\phi \phi}} = \frac {1} {r \sin\theta}$. $ \sigma^{i}$'s are
the Pauli matrices with $i=1,2,3$.

Our task is to find the solutions of eqn. (\ref{eq9}). First
substitute the wave function  eqn. (\ref{eq11})  and the matrices
eqn. (\ref{eq12}) in the generalized Dirac equation eqn.
(\ref{eq9}), and cancel the exponential factor. Since we are
working with WKB approximation, the contributions from $\partial
A$, $\partial B$ and high orders of $\hbar$ are neglected. We
finally obtain decoupled four Hamilton-Jacobi equations

\begin{eqnarray}
 - iA\frac{1}{\sqrt f }\partial _t I - B\left( {1 - \beta m^2}
 \right)\sqrt
g \partial _r I - Am\beta \left[ {g^{rr}\left( {\partial _r I}
\right)^2 + g^{\theta \theta }\left( {\partial _\theta I} \right)^2
+ g^{\phi \phi }\left( {\partial _\phi I} \right)^2} \right] \nonumber \\
 + B\beta \sqrt g \partial _r I \left[ {g^{rr}\left( {\partial _r I}
\right)^2 + g^{\theta \theta }\left( {\partial _\theta I}
\right)^2 + g^{\phi \phi }\left( {\partial _\phi I} \right)^2}
\right] + Am\left( {1 - \beta m^2} \right)  =  0, \label{eq13}
\end{eqnarray}

\begin{eqnarray}
iB\frac{1}{\sqrt f }\partial _t I - A\left( {1 - \beta m^2}
\right)\sqrt g
\partial _r I - Bm\beta \left[ {g^{rr}\left( {\partial _r I} \right)^2 +
g^{\theta \theta }\left( {\partial _\theta I} \right)^2 + g^{\phi
\phi }\left( {\partial _\phi I} \right)^2} \right] \nonumber \\
 + A\beta \sqrt g \partial _r I \left[ {g^{rr}\left( {\partial _r I}
\right)^2 + g^{\theta \theta }\left( {\partial _\theta I} \right)^2
+ g^{\phi \phi }\left( {\partial _\phi I} \right)^2} \right] + Bm\left( {1 -
\beta m^2} \right)
 =  0,
\label{eq14}
\end{eqnarray}

\begin{eqnarray}
A\{-(1-\beta m^2) \sqrt {g^{\theta \theta}}\partial _ {\theta} I +
\beta \sqrt {g^{\theta \theta}}\partial _ {\theta}I[g^{rr}
(\partial _r I)^2 + g^{\theta \theta}(\partial _ {\theta}I)^2
+ g^{\phi \phi}(\partial _ {\phi}I)^2 ]\nonumber\\
-i(1-\beta m^2) \sqrt {g^{\phi \phi}}\partial _ {\phi}I+ i\beta
\sqrt {g^{\phi \phi}}\partial _ {\phi}I[g^{rr} (\partial _r I)^2 +
g^{\theta \theta}(\partial _ {\theta}I)^2 + g^{\phi \phi}(\partial
_ {\phi}I)^2]\} = 0, \label{eq15}
\end{eqnarray}

\begin{eqnarray}
B\{-(1-\beta m^2) \sqrt {g^{\theta \theta}}\partial _ {\theta} I +
\beta \sqrt {g^{\theta \theta}}\partial _
{\theta}I[g^{rr}(\partial _r I)^2  + g^{\theta \theta}(\partial _
{\theta}I)^2
+ g^{\phi \phi}(\partial _ {\phi}I)^2 ]\nonumber\\
-i(1-\beta m^2) \sqrt {g^{\phi \phi}}\partial _ {\phi}I+ i\beta
\sqrt {g^{\phi \phi}}\partial _ {\phi}I[g^{rr} (\partial _r I)^2 +
g^{\theta \theta}(\partial _ {\theta}I)^2 + g^{\phi \phi}(\partial
_ {\phi}I)^2]\} = 0. \label{eq16}
\end{eqnarray}

\noindent To find the relevant solution, since the metric has a
time-like killing vector, we perform the separation of variables
as follows

\begin{equation}
I = - \omega t + W\left( r \right) + \Theta \left( \theta, \phi \right),
\label{eq17}
\end{equation}

\noindent where $\omega$ turns out to be the energy of the emitted
particle. We insert eqn. (\ref{eq17}) into eqn.s
(\ref{eq13})-(\ref{eq16}) and first focus on the last two
equations. They are identical after divided respectively by $A$
and $B$ and can be rewritten as follows

\begin{eqnarray}
(\sqrt {g^{\theta \theta}}\partial _ {\theta} \Theta +i\sqrt
{g^{\phi \phi}}\partial _ {\phi}\Theta)\nonumber\\
\times [\beta g^{rr}(\partial _r W)^2 + \beta g^{\theta
\theta}(\partial _ {\theta}\Theta)^2 +\beta g^{\phi \phi}(\partial
_ {\phi}\Theta)^2 - (1-\beta m^2)] =0. \label{eq18}
\end{eqnarray}

\noindent In the equation above, the value in the square bracket
can not vanish since $\beta$ is a small quantity representing the
effects from quantum gravity. Therefore, the expression in the
round brackets is zero and yields the solution of $\Theta $. In
the previous work, though $\Theta$ has a complex solution (other
than the trivial one $\Theta = constant$) and gives rise to a
contribution to the imaginary part of the action, it has no
contribution to the tunnelling rate. Therefore eqn. (\ref{eq18})
is simplified as

\begin{eqnarray}
\sqrt {g^{\theta \theta}}\partial _ {\theta} \Theta +i\sqrt
{g^{\phi  \phi}}\partial _ {\phi}\Theta =0. \label{eq19}
\end{eqnarray}

\noindent After cancelling $A$ and $B$, eqn. (\ref{eq13}) and
eqn. (\ref{eq14}) are identical and give rise to

\begin{equation}
 A_6\left( {\partial _r W} \right)^6 + A_4\left(
{\partial _r W} \right)^4 + A_2\left( {\partial _r W} \right)^2 + A_0 =
0,
\label{eq20}
\end{equation}

\noindent with
\begin{eqnarray}
A_6 & = & \beta^{2}g^{3}f,\nonumber \\
A_4 & = & \beta g^{2}f\left(m^{2}\beta+2\beta Q-2\right),\nonumber \\
A_2 & = & gf\left[\left(1-\beta m^{2}\right)^{2}+\beta\left(2m^{2}-
2m^{4}\beta-2Q+\beta Q^{2}\right)\right],\nonumber \\
A_0 & = & -m^{2}\left(1-\beta m^{2}-\beta Q\right)^{2}f-
\omega^{2},\nonumber \\
Q & = & g^{\theta\theta}\left(\partial_{\theta}\Theta\right)^{2}
+g^{\phi\phi}\left(\partial_{\phi}\Theta\right)^{2}.
\label{eq21}
\end{eqnarray}

\noindent Using eqn. (\ref{eq19}), we find $Q = 0$. Neglecting the
higher orders of $\beta$ and solving the above equations at the
event horizon yields \footnote{In \cite{Akhmedov:2006un}, the
authors argued that the invariance under canonical coordinate
transformation requires that the integral to calculate the
imaginary part of $W(r)$ should be a loop rather than an open
one-way integral. However, in our calculation, only the difference
between the imaginary parts matters.},

\begin{eqnarray}
W\left(r\right) & = & \pm\int\frac{1}{\sqrt{gf}}\sqrt{m^{2}
\left(1-2\beta m^{2}\right)f+\omega^{2}}\left(1+\beta
\left(m^{2}+\frac{\omega^{2}}{f}\right)\right)dr\nonumber \\
 & = & \pm i\pi2M\omega\left(1+\frac{1}{2}\beta
 \left(3m^{2}+4\omega^{2}\right)\right)+\left(\hbox{real part}\right).
\label{eq22}
\end{eqnarray}
\noindent In the above equation, $ f = g = 1-\frac{2M}{r}$. The
real part is irrelevant to the tunnelling rate. The $+ / -$ sign
corresponds to outgoing/ingoing wave. Then the tunnelling rate
\cite{Mitra} of the spin-1/2 fermion crossing the horizon is

\begin{eqnarray}
\Gamma & = &
\frac{P_{\rm(emission)}}{P_{\rm(absorption)}}=\frac{\exp
\left(-2\,\mathrm{Im} I_{+}\right)}{\exp\left(-2\,\mathrm{Im}
I_{-}\right)}=\frac{\exp\left(-2\,\mathrm{Im} W_{+}
-2\,\mathrm{Im} \Theta\right)}
{\exp\left(-2\,\mathrm{Im} W_{-}-2\,\mathrm{Im} \Theta\right)}\nonumber \\
& = & \exp\left[-8\pi M\omega\left(1+\frac{1}{2}\beta
\left(3m^{2}+4\omega^{2}\right)\right)\right].
 \label{eq23}
\end{eqnarray}

\noindent This is the Boltzmann factor with Hawking temperature

\begin{eqnarray}
T = \frac{1}{8\pi M\left(1+\frac{1}{2}\beta\left(3m^{2}+4\omega^{2}\right)\right)}
 = \left[1-\frac{1}{2}\beta\left(3m^{2}+4\omega^{2}\right)\right]T_{0},
\label{eq24}
\end{eqnarray}

\noindent where $T_0 = \frac{1}{8\pi M}$ is the original Hawking
temperature. It shows that there is a small correction to the
Hawking temperature, and the correction value is dependent not
only on the black hole's mass but also on the mass and energy of
emitted fermions. This property has been obtained in literature.
In \cite{PW}, energy conservation is enforced by dynamical
geometry and the tunnelling rate is found to be $\Gamma
=\exp[-8\pi \omega (M- \frac{\omega}{2})]$. Then the corrected
Hawking temperature is $T = \frac{1}{8\pi M-4\pi\omega}$, where
the leading correction to the Hawking temperature is related to
the energy of emitted particles. To address effects of quantum
gravity, the authors of \cite{NS,NHM} adopted the modified
commutation relation between the radial coordinate and the
conjugate momentum. They studied the quantum tunnelling of scalar
particles in the Schwarzschild black hole. The tunnelling rate was
derived as $\Gamma = \exp[-8\pi M\omega + 4\pi \omega^2 (3M a l_p
+1) - 8\pi \omega^3 (\frac{7}{3}M a^2 l^2_p +\frac{4}{3}a l_p) +
20\pi a^2 l^2_p\omega^4 +0(a^2 l^4_p)]$. Thus the correction to
the Hawking temperature is also related to the black hole's mass
and the particle's energy.

It is of interest to note that in eqn. (\ref{eq24}), the quantum
correction slows down the increase of the temperature during the
radiation. This correction therefore causes the radiation ceased
at some particular temperature, leaving the remnant mass. To
estimate the residue mass, it is enough to consider massless
particles. The temperature stops increasing when
\begin{equation}
(M-dM)(1+\beta \omega^2)\simeq M
\end{equation}
Then with the observation $dM = \omega$ and $\beta =\beta_0/M_p^2$
where $M_p$ is the Planck mass and $\beta_0 <10^{34}$ \cite{WYZ2}
is a dimensionless parameter marking quantum gravity effects, we
can get
\begin{equation}
M_{\hbox{Res}} \simeq \frac{M_p^2}{\beta_0 \omega} \gtrsim
\frac{M_p}{\beta_0}, \hspace{7mm} T_{\hbox{Res}} \lesssim
\frac{\beta_0}{8\pi M_p},
\end{equation}
where we have assumed the maximal energy of the radiated particle
is $\omega \simeq M_p$. This result is consistent with those
obtained in \cite{BG,Adler:2001vs, Chen:2002tu, Xiang:2006mg}.
Compared with previous results, our calculation explicitly shows
how the residue mass of black holes arises due to quantum gravity
effects. The singularity of black hole evaporation is then
prevented by the quantum gravity correction.

\section{Discussion and conclusion}

In this work, we modified the Dirac equation in curved spacetime
to  include the quantum gravity influence. To fulfill this
purpose, we employed the generalized uncertainty principle model.
This model is derived from the existence of minimal length which
arises when combine quantum and gravity. We calculated the
radiation of spin $1/2$ particles in the 4-dimensional
Schwarzschild spacetime with Hamilton-Jacob method. The tunnelling
rate and Hawking temperature were presented.

We found that the quantum gravity correction is related not only
to the black hole's mass but also to the mass (energy) of emitted
fermions. More interestingly, our result shows that the quantum
gravity correction explicitly retards the temperature rising in
the black hole evaporation. Therefore, at some point during the
evaporation, the quantum correction balances the traditional
temperature rising tendency. This leads to the existence of the
remnants. We showed that the remnants is $M_{\hbox{Res}}\gtrsim
\frac{M_p}{\beta_0}$, where $M_p$ is the Planck mass and $\beta_0
< 10^{34}$ from simple electroweak consideration. Therefore, the
classical thermodynamics singularity can be avoided and a residue
temperature $T_{\hbox{Res}} \lesssim \frac{\beta_0}{8\pi M_p}$ of
black holes exists.

We use the 4-dimensional Schwarzschild metric in this work. It is
of interest to employ other geometries in the studies. In our
calculation, we keep only the leading order of $\hbar$ and $\beta
= \beta_0/M_p^2$. It is expected that higher orders of corrections
may give more information in the future work.

%
\begin{acknowledgments}
D.Y. Chen and H. Wu are very grateful for Prof. S.Q. Wu and
Prof. P. Wang for their useful discussions. This work is supported
in part by the NSFC (Grant No. 11205125, 11175039, 11178018) and
SiChuan Province Science Foundation for Youths (Grant No.
2012JQ0039).
\end{acknowledgments}

\end{document}